\documentclass[prb,superscriptaddress,twocolumn,showpacs,floatfix]{revtex4-1}
\usepackage{amsmath}
\usepackage{amsfonts}
\usepackage{amssymb}
\usepackage{graphicx,graphics,color,epsfig}

\newcommand{\beq}{\begin{equation}}
\newcommand{\eeq}{\end{equation}}

\begin{document}

\title{Vertex functions at finite momentum: Application to antiferromagnetic
quantum criticality}
\author{Peter W\"{o}lfle}
\affiliation{Institute for Theory of
Condensed Matter and Institute for Nanotechnology, Karlsruhe Institute of Technology, 76049 Karlsruhe,
Germany}
\author{Elihu Abrahams}
\affiliation{Department of Physics and Astronomy, University of California
Los Angeles, Los Angeles, CA 90095} 

\date{\today{}}

\begin{abstract}
We analyze the three-point vertex function that describes the
coupling of fermionic particle-hole pairs in a metal to spin or charge fluctuations at non-zero momentum. We consider Ward identities, which connect two-particle
vertex functions to the self energy, in the framework of
a Hubbard model. These are derived using conservation laws following from local symmetries. The generators
considered are the spin density and particle density. It
is shown that at certain antiferromagnetic critical points, where the
quasiparticle effective mass is diverging, the vertex function describing
the coupling of particle-hole pairs to the spin density Fourier component at
the antiferromagnetic wavevector is also divergent. Then we give  an explicit calculation of the irreducible vertex function for the
case of three-dimensional antiferromagnetic fluctuations, and show that it is
proportional to the  diverging quasiparticle effective mass .
\end{abstract}

\maketitle

\preprint{}

\section{Introduction}

Over the past several decades the theoretical description of the
paramagnetic to antiferromagnetic (AFM) phase transition in metals has been
an ever-growing challenge (for a review see Ref.\ \onlinecite{lrvw07}). The theoretical description of systems for which the only critical degrees of freedom are
the (bosonic) spin fluctuations, while the fermionic quasiparticles (quasiparticles)
are not critical, is well-developed.\cite{moriya85,hertz76,millis93}\ However, in cases where the quasiparticles acquire
critical behavior, {\it e.g.} a diverging effective mass, as indicated by an
apparently diverging specific heat coefficient (prominent examples of compounds 
showing this behavior are CeCu$_{5.9}$Au$_{0.1}$  \cite{hvl96} and YbRh$_{2}$Si$_{2}$ \cite{geg02}), recent theoretical attempts\cite{metlitski10,lee13,lee14,efetov13} 
following the conventional field-theoretical methodology have not successfully explained experiment. 

Recently a semi-phenomenological theory of correlation functions $\chi (%
\mathbf{q},\nu)$ near an AFM quantum-critical point (QCP) has been proposed.\cite%
{wa11,aw12}. Since the prominent AFM fluctuations occur at a non-zero ordering wavevector $\mathbf{Q}$, the theory adopts a generalization,  to non-zero wavevector, of the usual Ward identity that relates the three-point spin-density vertex function $\Lambda({\bf q},\nu)$ in the limit of zero wavevector to the quasiparticle effective
mass.  Then, the usual weak-coupling form of the
dynamical spin response function $\chi (\mathbf{q},\nu )$ acquires
singular vertex corrections $\Lambda({\bf Q},\nu)$ to the Landau damping term; and the
coupling of bosons and fermions also gets enhanced by $\Lambda ({\bf Q})$.
Moreover, it was shown that spin exchange energy fluctuations at small wavevector (a fluctuation combining two spin fluctuation
propagators) are highly singular, and may lead to critical quasiparticles all over the
Fermi surface, not only at ``hot spots".\cite{asw14} A central result of the
theory \cite{wa11,aw12,asw14} is a self-consistent relation for the
quasiparticle effective mass, which allows for two very different solutions, depending on
initial conditions at high energy (or temperature): (1)  a weak-coupling solution similar to
the conventional SDW scenario \cite{moriya85,millis93} and (2) a strong-coupling solution characterized by critical quasiparticles with effective
mass that diverges as a fractional power law. The results of this theory are in
quite detailed agreement with experiments.\cite%
{hvl96,schroder00,trov00} It is therefore natural to ask how the
assumption of a singular vertex correction  as adopted in the phenomenological
theory \cite{wa11,aw12,asw14} may be derived from microscopic theory. In the
present paper we shall use two different approaches (Secs.~III and IV) to achieve that goal. 

In Sec.~II, we discuss generalized Ward identities, which  are based on conservation laws that follow from symmetries of the Hamiltonian and which are important in the context of a class of quantum-critical phenomena.  

In Sec.~III  we show how spin conservation leads to the required generalized Ward identity.


In Sec.\ IV,  we explicitly calculate the particle-hole irreducible
spin-density vertex function $\Lambda ({\bf Q})$ at non-zero $\mathbf{Q}$ near an
incommensurate antiferromagnetic QCP in the framework of the strong-coupling
theory developed in Refs.\ \onlinecite{wa11,aw12,asw14} and show that the vertex diverges like the effective mass, as assumed earlier in that theory. This
demonstrates that there is a closed self-consistent system of equations
connecting the two singular quantities, effective mass $m^{\ast }$ and
vertex function $\Lambda ({\bf Q}).$

\section{Symmetry, Ward identity and quantum criticality}

Symmetry properties are among the most important
pieces of knowledge characterizing a system. The standard symmetries, related
to invariance under translation in space and time, rotation in position
space, spin space or other internal spaces are well-known and are used to
develop methods of calculating the system's properties. In the realm of
quantum many-body physics and quantum field theory, symmetry properties may
be shown to give rise to useful relations among the two-particle and the one-particle Green's functions, the Ward-Takahashi identities.\cite{ward50, tak} These identities are usually derived by considering global symmetry
transformations effected by the application of unitary operators (gauge
transformations, rotations, etc.). Typically, a Ward identity relates the structure of the single-particle Green's function to a three-point vertex function $\Lambda({\bf q},\nu)$. A standard way \cite{nambu60} of constructing a Ward identity is to identify a conservation law that follows from a symmetry of the Hamiltonian and then using it to simplify the equation of motion for a two-particle Green's function that contains the three-point vertex.  Since we are primarily interested in an incommensurate antiferromagnetic QCP in the framework of the strong-coupling theory developed in Refs.  \onlinecite{wa11,aw12,asw14}, in what follows, we shall concentrate on the consequences of spin-rotational invariance.

The Ward identities are usually applied for the
limit of vanishing wavevector and frequency of an applied test field. They are therefore of limited use in characterizing fluctuations at non-zero wave
vector, such as antiferromagnetic fluctuations or charge density wave fluctuations in a metal.

However, the local conservation laws are valid on all spatial and temporal scales and give rise to generalized Ward identities even at non-zero wavevector. This has already been recognized and implemented by U. Behn, \cite{Behn} who used the procedure described above and which we elaborate in Sec.~III. Although these identities may be less stringent because, as we shall see, they could involve two
vertex functions, one of a density type, the other of a current-density type,
they nonetheless may be used to infer qualitative information. This is of
particular interest if single-particle properties, such as the
quasiparticle effective mass at the Fermi surface, are singular. This may
happen at a quantum critical point. In metallic compounds, quantum critical
points are often found to be of antiferromagnetic or charge-density wave
character, which involve fluctuations of spin or charge at non-zero wave
vector. Specific heat data in the neighborhood of such critical points often indicate a divergent quasiparticle
effective mass. Examples are many of the heavy-fermion compounds, some of
the iron-based superconductors and possibly the cuprate superconductors. The question becomes: How  does a singularity in the single-particle
properties affect the two-particle vertex functions at non-zero wavevector?
Here the generalized Ward identities may be useful. In the present paper, the answer will be given as
\beq
\Lambda({\bf Q},\nu \to 0)  \sim  Z^{-1}(p_+)+Z^{-1}(p_-),
\eeq
where $Z^{-1} = 1-\partial \Sigma/\partial \omega$ is proportional to the quasiparticle effective mass and $p_\pm = p \pm Q/2$ are the momenta of the incoming and outgoing legs of the three-point vertex  describing momentum transfer ${\bf Q}$ and energy transfer $\nu$


\section{Conservation law and continuity equation: Vertex function in linear response} 
\label{dg}

 We consider systems of identical fermions interacting via local
interactions. To be concrete, we  may take the Hubbard model, whose interaction term is of the form

\begin{equation}
{\cal H}_{\rm int}= U\int dr\Psi^{\dag}_\uparrow(r)\Psi_{\uparrow}(r)\Psi_{\downarrow}^{\dag}(r)\Psi_{\downarrow}(r)   
\label{hubbard}
\end{equation}%
where $\Psi_{\alpha}^{\dag }(r),\Psi_\alpha(r)$ creates or annihilates an electron
of spin $\alpha =\uparrow ,\downarrow$ at location r.

As a practical example, we consider the response of the system to an external magnetic field ${\vec H(r)}$ that couples to the spin density ${\vec \rho}_s(r) = \sum_{\alpha,\beta} \Psi^{\dag}_\alpha(r)
{\vec \sigma}_{\alpha\beta}\Psi_\beta(r)$ via an interaction term 

\beq
{\cal H}'= \lambda \int dr{\vec H(r)}\cdot{\vec \rho}_s(r).
\label{mag}
\eeq
Before showing how the spin-density vertex function, the expectation value of ${\vec \rho}_s(r,t)$ (in Heisenberg representation), controls the response to  ${\cal H}'$, we remind that when a density operator commutes with the interaction term in the Hamiltonian,  then it  obeys a continuity equation, derived from $\partial \Psi/\partial t = i[{\cal H},\Psi]$. Since the spin density operator commutes with  Eq.\ (\ref{hubbard}),  only the kinetic energy  enters the commutator and we have the  familiar local conservation law (repeated greek indices are summed): 

\beq
\frac{\partial {\vec \rho}_s(r,t)}{\partial t}- \frac{i}{2m}[{\vec\nabla}\cdot({\vec \nabla}\Psi^{\dag}_\alpha {\vec \sigma}_{\alpha\beta}\Psi_\beta-\Psi^\dag_\alpha {\vec\sigma}_{\alpha\beta}{\vec \nabla}\Psi_\beta]=0.
\label{cont}
\eeq%
The spatial derivatives in Eq.~(4) come from the commutator of $\rho_s$ with the kinetic energy, which in momentum space is $\sum_{k,\alpha}\epsilon_k c_{k,\alpha}^\dag c_{k,\alpha}$ For later convenience, it is useful to reexpress the continuity equation in momentum space:
\beq
i\partial {\vec \rho}_q /\partial t = \sum _k (\varepsilon_{kq}) c^\dag _{k_+,\alpha}{\vec \sigma}_{\alpha\beta}c_{k_-,\beta},
\eeq
where ${\vec \rho}_q = \sum_k c _{k_+,\alpha}^\dag{\vec\sigma}_{\alpha \beta}c_{k_-,\beta}$, with $k_\pm = k\pm q/2$
and $\varepsilon_{kq} =\epsilon_{k_+ }- \epsilon _{k_-}$
In compact form, it is
\beq
\sum_{k}D(k,q;t) \,c _{k_+\alpha}^\dag {\vec \sigma}_{\alpha\beta}c_{k_-,\beta}=0,
\label{cons}\eeq
 where the operator $D(k,q;t)$ is
 \beq
 D(k,q;t)= \frac{\partial}{\partial t} -(\varepsilon_{kq})
 \label{dkq}
 \eeq 
The considerations above apply also to the charge response and lead to the usual charge continuity equation. For our general purposes, we want to discuss the vertex function that describes the coupling of an electron (spin or charge) density to an external perturbation  such as  a charge or spin density fluctuation boson field or a magnetic field. For example, in linear response, the magnetic field perturbation ${\cal H}'$ of Eq.\ (\ref{mag})  gives rise to a correction to the single-particle Green's function $G_{\alpha\beta}(1,2) = -\langle T \Psi_\alpha(1)\Psi^{\dag}_\beta(2)\rangle,$ where $T$ is the time-ordering operator.
\beq
\delta G_{\alpha\beta}(1,2) = \int d3\ G^{(2)}_{\alpha\beta\delta\gamma}(1,2;3,3^+) \lambda {\vec H}(3)\cdot{\vec\sigma}_{\gamma\delta}.
\label{dg}
\eeq
 
Here, the two-particle Green's function is
\beq
G^{(2)}_{\alpha\beta\gamma\delta}(1,2;3,4) = -\langle T \Psi_\alpha(1)\Psi^\dag_\beta(2)\Psi_\gamma(3)\Psi^\dag_\delta(4)\rangle
\eeq	
and  in Eq.~(\ref{dg}), $3^+$ means $(r_3, t_3+0^+)$. It is seen that the RHS of Eq.\ (\ref{dg}) contains ${\vec \rho}_s(3)$; 
\beq
\delta G_{\alpha\beta}(1,2) = \int d3 \langle T \Psi_\alpha(1)\Psi^\dag_\beta(2){\vec \rho}_s(3)\rangle\cdot\lambda {\vec H}(3).
\label{2pgf}
\eeq
This shows how the vertex function controls the response.

Making use of the conservation law we may now derive identities relating the
response function to the single-particle Green's function or its
self-energy. To achieve this we re-express the  two-particle Green's function in Eq.~(\ref{2pgf}) in momentum space
and let the operator $D(p,q;t_3)$ that appears in the
conservation law, Eq.~(\ref{cons}) act on it. 
\begin{align}
G^{(2)} &\to  \langle T c_{k_+,\alpha}(t_1)c_{k_-,\beta}^{\dag}(t_2){\vec\rho}_q(t_3)\rangle \notag \\
&= \sum_p G^{(2)}_{\alpha\beta\delta\gamma}(k,p;q) {\vec \sigma}_{\gamma\delta},
\end{align}
where
\beq
G^{(2)}_{\alpha\beta\delta\gamma}(k,p;q) = - \langle T c_{k_+,\alpha}(t_1)c_{k_-,\beta}^\dag(t_2)c_{p_-,\delta}(t_3)c^\dag_{p_+\gamma}(t_3^+)\rangle.
\eeq
Here, $k_{\pm}= k\pm q/2, p_{\pm}= p \pm q/2$ are the momenta of the particle-hole pairs entering and leaving the two-particle Green's function
The action of the operator on the $\rho_q$ part of $G^{(2)}$ gives zero because of the conservation law, whereas the time derivative in $D$ acts on the step functions defined by the time ordering. The result is 
\beq
\sum_p[ \nu - \varepsilon_{pq}]G^{(2)}(k,p;q) = G(k_-) -G(k_+).
\label {res}
\eeq Here we have Fourier transformed in time $(t_3)$ and on the RHS of Eq.~(\ref{res}), $k$ and $q$ mean $({\bf k},\omega)$ and $({\bf q},\nu)$ respectively. That is, $k_\pm \to ({\bf k}\pm{\bf q}/2, \omega\pm\nu/2)$ and here the sum on $p$ includes $\int d\omega$. 
 For the charge response, Eq.~(\ref{res}) holds, as the spin indices are irrelevant. We shall restore them later, when necessary. 
 
 The next steps involve the use of well-known relations  among the Green's functions and associated amplitudes.
 \beq 
 G^{-1}({\bf k},\omega) = \omega-\epsilon_k -\Sigma{\bf k},\omega) = G_0^{-1}{\bf k},\omega) -\Sigma
 ({\bf k},\omega)
 \label{gf}
 \eeq
\beq
 G^{(2)}(k,p;q)= G(k_+)G(k_-)\Gamma(k,p;q)G(p_-)G(p_+)
\label{ggGgg}
\eeq

 \beq
 \Lambda(k,q) =1+ \sum_p\Gamma(k,p;q)G(p_-)G(p_+)
 \eeq
In the above, $\Gamma(k,p;q)$ is the four-point vertex (without external legs) and 
$\Lambda(k,q)$ is the three-point vertex amplitude that enters the Ward identity,
We use Eq.~(\ref{ggGgg}) in Eq.~(\ref{res}), divide out $G(k_+)G(k_-)$ and find
\begin{align}
&\sum_p(\nu - \varepsilon_{pq})\Gamma(k,p;q)G(p_-)G(p_+)\notag\\
&=  \nu - \varepsilon_{kq}
-\Sigma(k_+)+\Sigma(k_-),
\label{fin}
\end{align}
where $\Sigma(k)$ is the self energy part as in Eq.~(\ref{gf}). This result has already been anticipated in Ref.~\onlinecite{Behn}; we make use of it in what follows.

We take the derivative of  Eq.~(\ref{fin}) with respect to $\nu$ and as we are interested in the behavior of $\Lambda({\bf k},\omega;{\bf q},\nu)$ for $\nu\to 0$ and arbitrary ${\bf q}$, we then take $\nu= 0$ and obtain
\begin{align}
\Lambda(k;{\bf q},0) &-\sum_p\varepsilon_{pq}[\frac{\partial}{\partial \nu}\Gamma(k,p;q)G(p_-)G(p_+)]_{\nu=0}\notag \\
&= \frac{1}{2}[Z^{-1}({\bf p}_+,\omega) + Z^{-1}({\bf p}_-,\omega)],
\label{Z+Z}
\end{align}
where we used the quasiparticle weight factor at $({\bf p},\omega)$  defined as
\beq
Z^{-1}({\bf p},\omega) = 1 - \frac{\partial}{\partial \omega}\Sigma({\bf p},\omega)
\eeq

The second term on the LHS of the key result Eq.~(\ref{Z+Z}) is the $\nu=0$ derivative of the spin-current density vertex.

The spin (and charge) continuity equations, as in Eq.~(\ref{cont}) are a consequence of invariance under certain unitary transformations ({\it e.g.} gauge, rotation). However, as emphasized by Nambu,\cite{nambu60} there may also be non-unitary transformations under which the local spin and charge densities and/or the Hamiltonian are invariant (perhaps up to a shift in the chemical potential $\mu_0$). These can lead to ``pseudo-conservation laws"  that are operator identities for charge and spin density and current and to Ward-like relations similar to Eq.~(\ref{Z+Z}). \cite{pw} For example, carrying out the steps above, using the pseudo-conservation law,
\begin{align}
&\sum _k{\vec \sigma}_{\alpha\beta}\big\{[i\frac{\partial c^\dag _{k_+,\alpha}}{\partial t }c_{k_-,\beta} - ic^\dag _{k_+,\alpha} \frac{\partial c_{k_-,\beta}}{\partial t}] \notag\\ 
&+ [\varepsilon_{k_+}+\varepsilon_{k_-}  -2\mu_0] c^\dag _{k_+,\alpha}c_{k_-,\beta}\big\}=0,
\end{align}
 one may derive
\begin{align}
& \Lambda(k;\mathbf{q},0)+\frac{1}{2}\sum_{p'}
[2\omega ^{\prime }-\epsilon _{\mathbf{p}_{+}^{\prime }}-\epsilon _{%
\mathbf{p}_{-}^{\prime }}+2\mu _{0}\mathbf{]}\frac{\partial }{\partial \mu
_0}\Lambda(p,p';\mathbf{q},0)  \notag \\
& =\frac{1}{2}[Z^{-1}_{\mu}(\mathbf{p}_{+},\omega )+Z^{-1}_{\mu}(\mathbf{p}_{-},\omega )]
\label{WI_mu}
\end{align}
where $Z^{-1}_{\mu}(\mathbf{p},\omega )=1-\partial\Sigma({\bf p},\omega)/\partial\mu
_0$. 

Now, a situation of particular interest arises if the quasiparticle weight factor 
$Z(\mathbf{p},\omega )$ happens to be small, or even tends to vanish,
implying that the effective quasiparticle mass $m^{\ast }/m=Z^{-1}(\mathbf{p},\omega )$ is large either in certain regions on the
Fermi surface (so-called hot spots) or all over the Fermi surface. This will be
the case in the critical regime near a quantum phase transition  to {\it e.g.} an
antiferromagnetic phase. We may then conclude from  the key equation (\ref{Z+Z}) that
the three-point vertex is enhanced approximately proportional to the effective mass
enhancement. This follows from the fact that both the spin density and the
spin-current density vertices are given by integrals of $ \Gamma G(p_+)G(p_-)$ multiplied by two
different weight factors, $1$ and $\varepsilon _{pq}(\partial /\partial\nu)$, respectively.  Although the effective mass
enhancement occurs for a state near the Fermi surface, we emphasize the new result that
the vertex enhancement takes place if at least one of the partners of the
particle-hole pair, with momenta $\mathbf{p}_{+}$ or $\mathbf{p}_{-}$ , is
on the Fermi surface.
In order to demonstrate that Eq.~(\ref{Z+Z}) does indeed imply a proportionality of $\Lambda$ to $Z^{-1}$ we consider the limit of small, but finite $q$ and $\nu=0$, when Fermi liquid theory applies. In this case the vertex function is given as $\Lambda=Z^{-1}/(1+F_{a})$, where $F_{a}$ is the Landau parameter in the spin channel. In order for the Ward identity Eq.~(\ref{Z+Z}) to be satisfied, the current density term has to amount to $Z^{-1}F_a/(1+F_{a})$. The two contributions add up to $Z^{-1}$, as required by the Ward identity. In other words, for any non-zero Fermi liquid interaction $F_a$ both terms on the l.h.s. of Eq.~(\ref{Z+Z}) are proportional to $Z^{-1}$, and will therefore diverge whenever $Z^{-1}$ diverges.

\section{Vertex function at large Q}

We now calculate the irreducible spin-density vertex function $\Lambda (%
\mathbf{k},\omega ;\mathbf{Q},\nu =0)$ (called $\lambda _{Q}$ in Ref.~\onlinecite%
{asw14}) \ in the framework of the theory of critical quasiparticles near an antiferromagnetic critical point as developed in Refs.~\onlinecite{wa11,aw12,asw14}.
There it was shown that for the case of three-dimensional
AFM spin fluctuations, when conventional spin density wave theory is
supposed to work, a new strong-coupling regime may be accessible under
certain conditions. This regime is characterized by a power-law divergence
of the effective mass as a function of energy, and hyperscaling with
critical exponents $z=4$ and $\nu =1/3$. The theory requires the
particle-hole irreducible spin density vertex function at wavevector $%
\mathbf{Q}$ to diverge like the effective mass. As shown in Sec.~III, this
can be a consequence of the Ward identities. However, since the
Ward identities relate the full vertex functions $\Lambda $\ to the
effective mass (or the inverse quasiparticle weight factor $1/Z$), one may
ask how the irreducible vertex, which is the quantity needed in the strong
coupling theory \cite{wa11,aw12,asw14} depends on $Z$. We therefore show in
the following that a certain diagram contributing to the irreducible vertex
correction is indeed proportional to $1/Z$, provided one assumes that this
very vertex correction renormalizes the spectrum of spin fluctuations and
their the coupling to quasiparticles in just the way that was assumed in the theory of critical quasiparticles.

There are two ways in which the vertex function $\lambda _{Q}$ enters the
theory: first, the spin-fluctuation spectrum is affected in the Landau
damping term; it acquires a factor $\lambda _{Q}^{2}$,  from the renormalization of the
particle-hole bubble diagram  of Landau damping at each end. Thus
\begin{equation}
{\rm Im}\chi (\mathbf{q},\nu )=\frac{N_0\lambda _{Q}^{2}\nu }{(r+(\mathbf{%
q-Q})^{2})^{2}+(\lambda _{Q}^{2}\nu )^{2}}  \notag
\end{equation}%
Here $N_{0}$ is the bare density of states, $r$ is the dimensionless tuning
parameter ($r\rightarrow 0$ at the QCP) and wavevector $\mathbf{q}$\textbf{\ 
}and frequency\textbf{\ }$\nu $ are in units of $k_{F}$ and $\epsilon _{F}$,
respectively.\ Second, since the coupling of the spin fluctuations to the
quasiparticles also involves a factor $\lambda _{Q}$, each end
of a spin fluctuation line  receives a factor $\lambda _{Q}N_{0}^{-1}$.

The large momentum transfer involved in a scattering process of
quasiparticles off AFM spin fluctuations usually takes quasiparticles into final states
far from the Fermi surface, except for momenta at ``hot spots" on the Fermi
surface. The consequences of these limitations of critical scattering are
often not compatible with what is observed experimentally. It was therefore suggested In Ref.~\onlinecite{asw14} that simultaneous scattering off two spin
fluctuations with opposite momenta, leading to small total momentum transfer
would be a more relevant process.\cite{subir} Two spin fluctuations may be thought of as
an (exchange) energy fluctuation $\chi _{E}(\mathbf{q},\nu )$. The
corresponding spectrum was calculated in Ref.~\onlinecite{asw14}, Eq.~(3) to be

\begin{equation}
{\rm Im}\ \chi _{E}(\mathbf{q},\nu )=\frac{(N_{0})^{3}\lambda _{Q}^{3}\nu
^{5/2}}{(r+q^{2})^{2}+(\lambda _{Q}^{2}\nu )^{2}}  
\end{equation}
The corresponding self energy due to energy fluctuation exchange is given by 
\beq
\Sigma({\bf k},\omega) \sim \int dq\,G({\bf k}+{\bf q}, \omega  +\nu)\ \chi _{E}(\mathbf{q},\nu )
\label{sig}
\eeq
and leads to $\Sigma\propto \omega^{3/4}$ (and hence $Z(\omega)\propto\omega^{1/4}$)

The first vertex correction diagram that corresponds to the dressing of the spin-density vertex $\lambda_Q$ by energy fluctuations has one energy fluctuation that bridges the vertex:
\beq
\lambda _Q^{(1)}=\lambda_Q^4 \lambda _v^2 u^4\int dq\,G(p+q+Q)G(p+q)\chi _E(q)  
\eeq
Here $\lambda _{v}\propto 1/Z$ is a vertex correction at small momentum $%
\mathbf{q}$ governed by the usual Ward identity at ($\mathbf{q}=0,\nu
\rightarrow 0)$ and $u\propto N_{0}^{-1}$.
For generic $\mathbf{p}$ one of
the momenta, $\mathbf{p+q+Q}$ will be far from the Fermi momentum, while the other, $\mathbf{p+q%
}$, is close to it (or vice versa). We may then put $G(p+q+Q)\approx 1/\epsilon_F$. What remains is the self-energy expression, Eq.~(\ref{sig}), so
that 
\beq
\lambda _{Q}^{(1)}(\mathbf{p,}\omega ;\mathbf{Q},\nu =0)\approx \frac{%
\Sigma (\mathbf{p,}\omega )}{\epsilon _{F}}
\eeq%
We see that $\lambda _{Q}^{(1)}\to 0$ as $\omega \to 0$ and is not singular. 
However,  singular diagrams do occur if at least three  spin fluctuation lines in parallel are internal in a contribution to $\lambda _{Q}(\mathbf{p,}\omega ;\mathbf{Q},\nu=0)$ (any odd number will do). Two of these combine into an energy fluctuation. The resulting diagram has a spin fluctuation and an energy fluctuation
in the intermediate state, similar to the Azlamasov-Larkin diagram in the theory of superconducting fluctuations.
\beq
\lambda _{Q}^{(3)}=A\int dq\,G(p-q)T(q;Q)\chi (Q+q)\chi _{E}(q), 
\notag
\end{equation}
where $A= \lambda
_{Q}^{6}\lambda _{v}^{2}u^{6}$ and we defined the triangle loop
\begin{equation}
T(q;Q)=\int dp'\,G(p^{\prime }+q)G(p^{\prime })G(p^{\prime }+q+Q) 
\notag
\end{equation}%
The quantity $T(q;Q)$ is noncritical and may be replaced by $T(q;Q)\approx
N_{0}/\epsilon _{F}$ . It is convenient to first calculate the imaginary
part of $\lambda _{Q}^{(3)}$ at temperature $T<<\omega$:
\beq
{\rm Im}\lambda _Q^{(3)}
\approx A_1\int_{0}^{\omega }d\nu \int d{\vec q}\ {\rm Im}\chi (Q+q){\rm Im}\chi
_{E}(q) {\rm Im}G(p-q)
\eeq
Here $
A_1 = \lambda _{Q}^{6}\lambda _v^2 u^6(N_0/\epsilon _F)$.
The result of the integration over  the solid angle of ${\vec q}$ is $\propto 1/q$. We restrict ourselves to the critical
regime, $r=0$ in $\chi_E$. The $q$-integration may then be performed for $\lambda
_{Q}^{2}|\nu |<q^{2}<\infty $

\begin{align}
\rm{Im}\lambda _{Q}^{(3)}
&\propto \frac{k_{F}^{3}}{N_{0}\epsilon _{F}}\lambda _{Q}^{11}\lambda
_{v}^{2}\int_{0}^{\omega }d\nu |\nu |^{7/2}\int qdq\frac{1}{%
[q^{4}+(\lambda _{Q}^{2}|\nu |)^{2}]^{2}}  \notag \\
& \propto \lambda _{Q}^{5}Z^{-2}|\omega |^{3/2}  \notag
\end{align}%
where we used $\lambda _{v}\propto Z^{-1}$ as stated above.
We now identify $\lambda _{Q}^{(3)}=\lambda _{Q}$\ , 
and solve the resulting equation for $\lambda _{Q}$:

\begin{equation}
\lambda _{Q}\propto Z^{1/2}|\omega |^{-3/8}  
\label{one}
\end{equation}
This result may be combined with the result for $Z$ in the strong-coupling regime which was obtained in Ref.~\onlinecite{asw14}, Eq.~(4):

\begin{equation}
Z\propto \lambda _{Q}^{5}|\omega |^{3/2} 
\label{2}
\end{equation}%
Combining Eqs.~(\ref {one},\ref{2}), we find

\begin{equation}
\lambda _{Q}\propto Z^{-1}\propto |\omega |^{-1/4},  \notag
\end{equation}%
which is precisely what has been postulated in Ref.~\onlinecite{asw14} on the basis of
phenomenological arguments.

\section{Conclusion}

The Ward-Takahashi identities are based on conservation laws; in interacting electron systems they relate the single-(quasi)particle renormalizations to vertex amplitudes that describe response to external probes and/or coupling to collective modes. While these identities are usually given, both in quantum field theory and in the many-body problem, for the limit of momentum $\vec{q} \to 0$ and frequency $\nu \to 0$, there are cases for which $\vec{q} \neq 0$ is of interest, as when quasiparticles interact with non-uniform external fields or collective fluctuations.
Of particular interest is the situation in the neighborhood of a quantum critical point associated with an ordered phase having  a non-uniform order parameter. Examples are an antiferromagnetic or charge density wave state with a non-zero ordering vector. Here, quasiparticles couple to nonzero $\vec{q}$ order parameter fluctuations and the quasiparticle renormalization may be singular. The generalized Ward identity shows how the vertex amplitude for this coupling also acquires singular behavior, following the renormalized quasiparticle mass. A new feature is that it is sufficient that (at least) one of the  external lines to the (three-point) vertex is on the Fermi surface, rather than requiring both to be. Therefore, the identity holds  even when $\vec{q}$ does not connect two points on the Fermi surface (``hot spots"). Our results are of use in analyzing behavior of metals near quantum critical points.

\section{Acknowledgements}

We acknowledge useful discussions with A. V. Chubukov, G. Kotliar, D. Maslov, Q. Si, C. M. Varma, and especially J. Schmalian.  P.W. thanks the Department
of Physics at the University of Wisconsin-Madison for hospitality during several
stays as a visiting professor and acknowledges an ICAM senior scientist
fellowship. Part of this work was performed during the summers of 2012-14 at
the Aspen Center for Physics, which is supported by NSF Grant No.
PHY-1066293. P.W. acknowledges financial support by the Deutsche
Forschungsgemeinschaft through Grant No. SCHM 1031/4-1 and by the research unit FOR960 `Quantum Phase Transitions'.

\end{document}